\documentclass[preprint,12pt]{elsarticle}

\usepackage{graphicx}
\usepackage{color}
\usepackage[colorlinks,bookmarks=false,citecolor=blue,linkcolor=red,urlcolor=blue]{hyperref}

\usepackage{amssymb,amsbsy}
\usepackage{amsmath}

\newcommand{\op}[1]{%
    \fontdimen12\textfont3=2pt\fontdimen12\scriptfont3=1.4pt%
    \!\null\mathop{\vphantom{#1}\smash{#1}}\limits_{\sim}\null\!}

\newcommand{\ddt}{\frac{d}{dt}\;}
\newcommand{\fmref}[1]{(\protect\ref{#1})}

\newcommand{\pp}[2]{\frac{\partial \, {#1}}{\partial \, {#2}}\;}
\newcommand{\figref}[1]{Fig.~\protect\ref{#1}}

\journal{J. Magn. Magn. Mater.}

\begin{document}

\begin{frontmatter}



\title{Rotational magnetocaloric effect of anisotropic giant spin systems}


\author{Christian Beckmann\fnref{bi}}
\author{Julian Ehrens\fnref{bi}}
\author{J{\"u}rgen Schnack\corref{cor1}\fnref{bi}}
\ead{jschnack@uni-bielefeld.de}
\cortext[cor1]{corresponding author}
\address[bi]{Dept. of Physics, Bielefeld University, P.O. box
  100131, D-33501 Bielefeld, Germany}

\begin{abstract}
The magnetocaloric effect, that consists of adiabatic
temperature changes in a varying external magnetic field,
appears not only when the amplitude is changed, but in cases of anisotropic
magnetic materials also when the direction is varied. In this
article we investigate the magnetocaloric effect theoretically
for the archetypical single molecule magnets Fe$_8$ and
Mn$_{12}$ that are rotated with respect to a magnetic field. We
complement our calculations for equilibrium situations with
investigations of the influence of non-equilibrium thermodynamic
cycles. 
\end{abstract}


\end{frontmatter}


\section{Introduction}
\label{sec-1}

Magnetocalorics is an important thermodynamic concept with many
applications for instance in room-temperature or sub-kelvin
cooling
\cite{GiM:PR33,PeG:JMMM99,WKS:PRL02,ELJ:JMC06,GGC:RSER13,SCM:NC14,BRG:CM16}. It
rests 
to a large extend on the magnetocaloric effect (MCE) \cite{Smi:EPJH13}
which states that in adiabatic processes, i.e. processes with
constant entropy, the temperature changes upon the variation of
the external magnetic field. Nowadays's research efforts focus
e.g. on new materials \cite{TZC:APL10,SanSM12,BML:npjQM18} or theoretical optimization
of (molecular) magnetic materials
\cite{EvB:DT10,GCS:13,ELP:APL14,GCS:APL14}. 

While the typical magnetocaloric process employs variations of
the magnitude of the external field, some ideas focus on the
effect of a rotation of the field or equivalently of the sample
\cite{LRE:ANIE16,KPC:IC17,THB:APL00,ZWZ:PRL01}. A practical
reason for this approach is given by the fact that mechanical
rotations of the sample can be performed much more quickly than
field sweeps. However, the rotational magnetocaloric effect (rMCE)
requires anisotropic magnetic materials \cite{TTO:PB17,TTO:PB18}. 
In this article we
therefore discuss how single-molecule magnets (SMM) perform 
as cooling agents in the kelvin temperature region. These
materials are usually known for two other 
effects: slow relaxation and quantum tunneling of the magnetization
\cite{SGC:Nat93,FST:PRL96,TLB:Nature96,WeS:Science99} and not considered
for magnetic cooling in the ordinary sense. Nevertheless, their
large anisotropy makes them prospective candidates for the rMCE
\cite{THB:APL00}. 

In this article we investigate two aspects of the rMCE. We first
discuss the isentropes and isothermal entropy changes that can
be achieved in SMMs such as Fe$_8$ and Mn$_{12}$. In a second
part we set up a simple relaxation dynamics in order to estimate
how more realistic time-dependent Carnot processes would perform 
for these molecular coolers.

The article is organized as follows. In section~\ref{sec-2} we
introduce the model. Section~\ref{sec-3} deals with the
equilibrium rotational magnetocaloric effect whereas
section~\ref{sec-4} discusses dynamical aspects. Our results are
summarized in section~\ref{sec-5}.

\section{Model and numerical procedures}
\label{sec-2}

\subsection{Model Hamiltonians}

The low-temperature properties of single-molecule magnets with a
large spectral gap between the ground state multiplet and
higher-lying states can be rationalized using the giant spin
approximation. To this end the zero-field split ground state
multiplet is generated by an effective one-spin Hamiltonian. 
For Fe$_8$ the following giant spin Hamiltonian
was developed \cite{WeS:Science99}
\begin{eqnarray}
\label{E-2-1}
\op{H}_{\text{Fe$_8$}}
&=&
D_{\text{Fe$_8$}}\;
\op{S}_z^2
+
E_{\text{Fe$_8$}}\;
\left(
\op{S}_x^2 - \op{S}_y^2
\right)
+
g_{\text{Fe$_8$}}\, \mu_B\, \vec{B}(t)\cdot
\op{\vec{S}}
\\
&&
+
B_{4,\text{Fe$_8$}}^{0}\;
\op{O}_{4}^{0}
+
B_{4,\text{Fe$_8$}}^{2}\;
\op{O}_{4}^{2} 
+
B_{4,\text{Fe$_8$}}^{4}\;
\op{O}_{4}^{4}
\nonumber
\ ,
\end{eqnarray}
with specific values of $S=10$, 
$D_{\text{Fe$_8$}}=-0.295$~K, $E_{\text{Fe$_8$}}=0.05605$~K, 
$|E_{\text{Fe$_8$}}/D_{\text{Fe$_8$}}|=0.19$,
$B_{4,\text{Fe$_8$}}^{0}=2.3\cdot 10^{-6}$~K, 
$B_{4,\text{Fe$_8$}}^{2}=-7.2\cdot 10^{-6}$~K, 
$B_{4,\text{Fe$_8$}}^{4}=-1.2\cdot 10^{-5}$~K,
and $g_{\text{Fe$_8$}}=2.0$ \cite{BGS:CAEJ00}.

In the case of Mn$_{12}$, Hamiltonian \fmref{E-2-2} proved to be
appropriate 
\begin{eqnarray}
\label{E-2-2}
\op{H}_{\text{Mn$_{12}$}}
&=&
D_{\text{Mn$_{12}$}}\;
\op{S}_z^2
+
g_{\text{Mn$_{12}$}}\, \mu_B\, \vec{B}(t)\cdot
\op{\vec{S}}
\\
&&
+
B_{4,\text{Mn$_{12}$}}^{0}\;
\op{O}_{4}^{0}
+
B_{4,\text{Mn$_{12}$}}^{4}\;
\op{O}_{4}^{4}
\nonumber
\ ,
\end{eqnarray}
with specific values of $S=10$, 
$D_{\text{Mn$_{12}$}}=-0.65$~K, 
$B_{4,\text{Mn$_{12}$}}^{0}=-3.0\cdot 10^{-5}$~K, 
$B_{4,\text{Mn$_{12}$}}^{4}=\pm 4.6\cdot 10^{-5}$~K,
and $g_{\text{Mn$_{12}$}}=2.0$
\cite{BKH:JLTP05}.
Higher order spin operators are expressed by means of Stevens
operators 
\begin{eqnarray}
\label{E-2-3}
\op{O}_{4}^{0}
&=&
35\op{S}_z^4 - \left[30 S \left(S+1\right)-25\right]\op{S}_z^2 +
3 S^2\left(S+1\right)^2 - 6 S \left(S+1\right)
\\
\op{O}_{4}^{2}
&=&
\frac{1}{4} \left[7\op{S}_z^2 - S \left(S+1\right) 
- 5\right]\left((\op{S}^+)^2 +  (\op{S}^-)^2 \right) 
\\
&&+ 
\frac{1}{4}\left((\op{S}^+)^2 +  (\op{S}^-)^2
\right)\left[7\op{S}_z^2 
- S \left(S+1\right) - 5\right]
\nonumber
\\
\op{O}_{4}^{4}
&=&
\frac{1}{2}\left((\op{S}^+)^4 +
(\op{S}^-)^4\right)
\ .
\end{eqnarray}
It turns out that for the magnetocaloric investigations of this
article only the terms given in the respective first lines of
eqs.~\fmref{E-2-1} and \fmref{E-2-2} are relevant. The other
terms, which determine the tunnel splitting, have a stark effect
on the magnetization tunneling, but not on the thermal 
properties for the temperature and field ranges as well as
orientations studied here. We therefore use only the first lines
of \fmref{E-2-1} and \fmref{E-2-2} in the following
calculations.

\begin{figure}[ht!]
\centering
\includegraphics[width=0.2\textwidth]{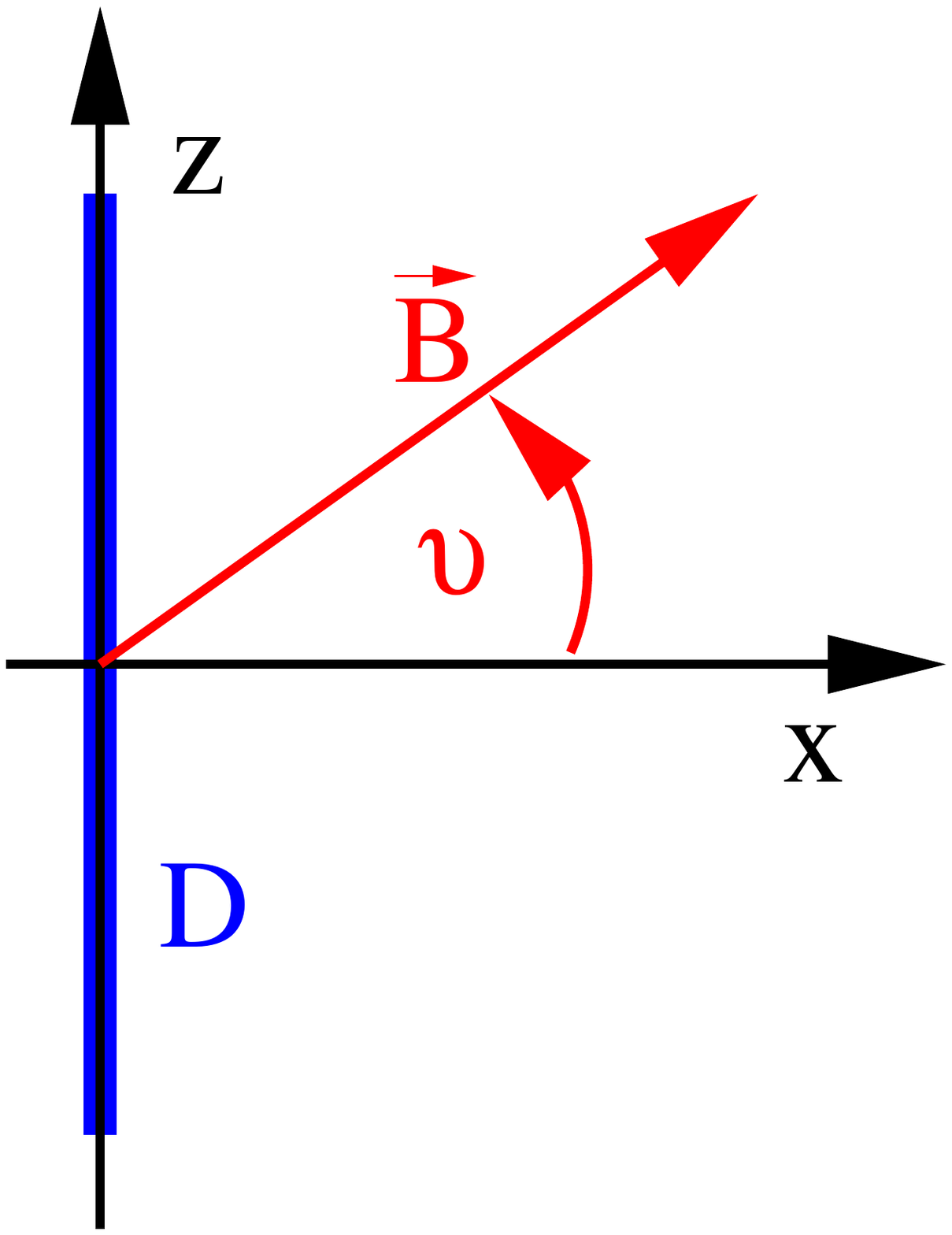}
\caption{Coordinate system used in the simulations.\label{jmmm-amce-f-a}}
\end{figure}
Since we are going to investigate the influence on the relative
orientation of the magnetic field with respect to the easy axis
of the SMM, we define the coordinate system as shown in
Fig.~\ref{jmmm-amce-f-a}. The magnetic easy axis is parallel to
the $z$-axis, whereas the field vector lies in the $xz$-plane
with an inclination of $\vartheta$ with the positive $x$-axis.

\subsection{Equilibrium thermodynamics}

Here and throughout the article we tacitly assume that the
applied magnetic fields and the investigated temperatures do not
violate the ranges of applicability of the giant spin models 
\fmref{E-2-1} and \fmref{E-2-2}.

Equilibrium thermodynamic observables can be obtained from
the Gibbs potential $G(T,\vec{B})$
\begin{eqnarray}
\label{E-2-4}
G(T,\vec{B})
&=&
-k_B T \
\text{log}\left[Z(T,\vec{B})\right]
\ ,
\end{eqnarray}
where $Z(T,\vec{B})$ denotes the partition function. Entropy as
well as magnetization are first derivatives of $G(T,\vec{B})$,
i.e. 
\begin{eqnarray}
\label{E-2-5-S}
{\mathcal S}(T,\vec{B})
&=&
-
\pp{}{T}\, G(T,\vec{B})
\ ,
\\
\label{E-3-5-M}
\vec{{\mathcal M}}(T,\vec{B})
&=&
-
\pp{}{\vec{B}}\, G(T,\vec{B})
\ .
\end{eqnarray}

\subsection{Non-equilibrium thermodynamics}

The infinitesimal work $\delta W$ is defined by the variation of
the magnetization 
\begin{eqnarray}
\label{E-3-1}
\delta W & = & \vec{B}\cdot\text{d}\vec{\mathcal{M}}\ .
\end{eqnarray}
Since this term is complicated to evaluate we use
the following relation: 
\begin{eqnarray}
\oint\text{d}\left(\vec{\mathcal{M}}\cdot\vec{B}\right)=0 & = &
\oint\vec{B}\cdot\text{d}\vec{\mathcal{M}}+\oint\vec{\mathcal{M}}\cdot\text{d}\vec{B}\\ 
\Rightarrow\;\Delta W & = & -\oint\vec{\mathcal{M}}\cdot\text{d}\vec{B}\:.\label{E-3-2}
\end{eqnarray}
The magnetization $\vec{\mathcal{M}}$ can be obtained from the time-dependent
density matrix. 
For the time evolution of the density matrix we employ
\begin{eqnarray}
\ddt\op{\rho}\left(t\right) & = &
-i\left[\op{H},\op{\rho}\left(t\right)\right]
+c\cdot\lambda\left(\op{\rho}^{(\text{eq})}(T,\vec{B})-\op{\rho}(t)\right)\ ,\label{E-3-3}
\end{eqnarray}
with 
\begin{eqnarray}
\op{\rho}^{(\text{eq})}(T,\vec{B}) 
& = & \frac{1}{Z(T,\vec{B})}\sum_{n}\left|n\right\rangle e^{-\beta E_{n}}\left\langle n\right|\ ,\label{E-3-4}
\end{eqnarray}
where $\beta=(k_{B}T)^{-1}$. $E_{n}$ and $\left|n\right\rangle $
denote the eigenvalues and the corresponding eigenvectors of the Hamiltonian.
The factor $c$ depends on the nature of the current stroke in
the process. For an isothermal process $c=1$, for an adiabatic
(or isolated) process $c=0$. The factor $\lambda$ denotes the
coupling strength of the system to one of the heat reservoirs and
can in principle be different for each heat bath.

The mean magnetization can easily be calculated from the density matrix:
\begin{eqnarray}
\vec{\mathcal{M}} \left(t\right) 
& = & -g\,\mu_{B}\,\text{Tr}\left\{ \op{\vec{S}}\op{\rho}\left(t\right)\right\} \;.\label{E-3-5}
\end{eqnarray}
For the mean work $\Delta W$ done on or by the system during
one stroke in the time interval $\left[t_{0}\:,\:t_{1}\right]$ one
finds with \fmref{E-3-2} 
\begin{eqnarray}
\Delta W & =- & \int_{t_{0}}^{t_{1}}
\vec{\mathcal{M}}\left(t\right) \cdot \dot{\vec{B}}\left(t\right)\text{d}t
\ .
\label{E-3-6}
\end{eqnarray}
The during this process absorbed or emitted amount of heat $\Delta Q$
can be obtained from the change of the mean energy $\left\langle \op{H}\right\rangle $
of the system via the rules of thermodynamics: 
\begin{eqnarray}
\left\langle \op{H}\right\rangle \left(t\right) 
& = & \text{Tr}\left\{ \op{H}\left(t\right)\op{\rho}\left(t\right)\right\} \ ,\\
\Delta\left\langle \op{H}\right\rangle  
& = & \left\langle \op{H}\right\rangle
\left(t_{1}\right)-\left\langle \op{H}\right\rangle
\left(t_{0}\right)\ 
=\;\Delta Q+\Delta W\;.\label{E-3-7}
\end{eqnarray}

\begin{figure}[ht!]
\centering
\begin{tabular}{cc}
  \includegraphics[width=0.3\textwidth]{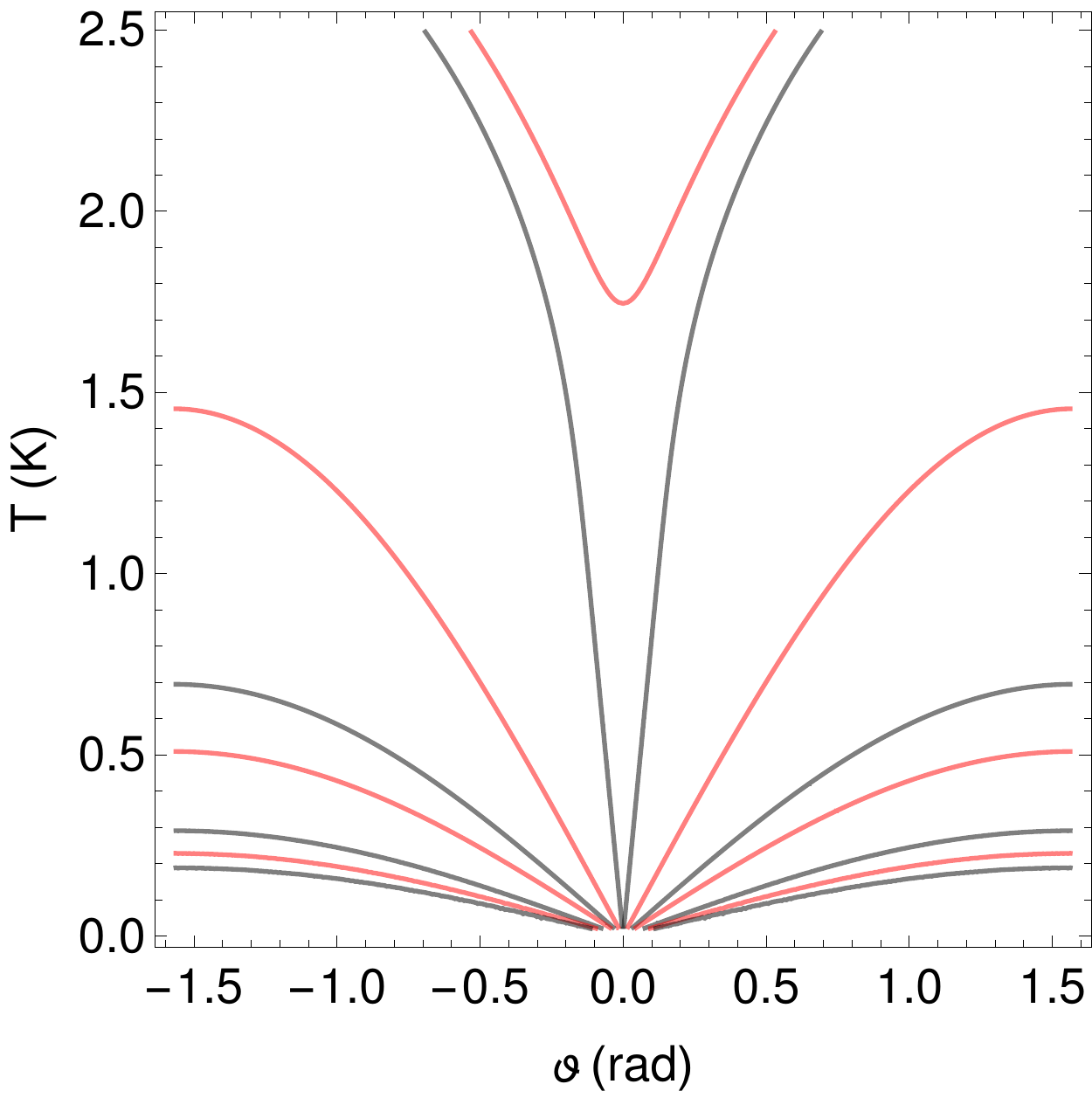}
  &
  \includegraphics[width=0.3\textwidth]{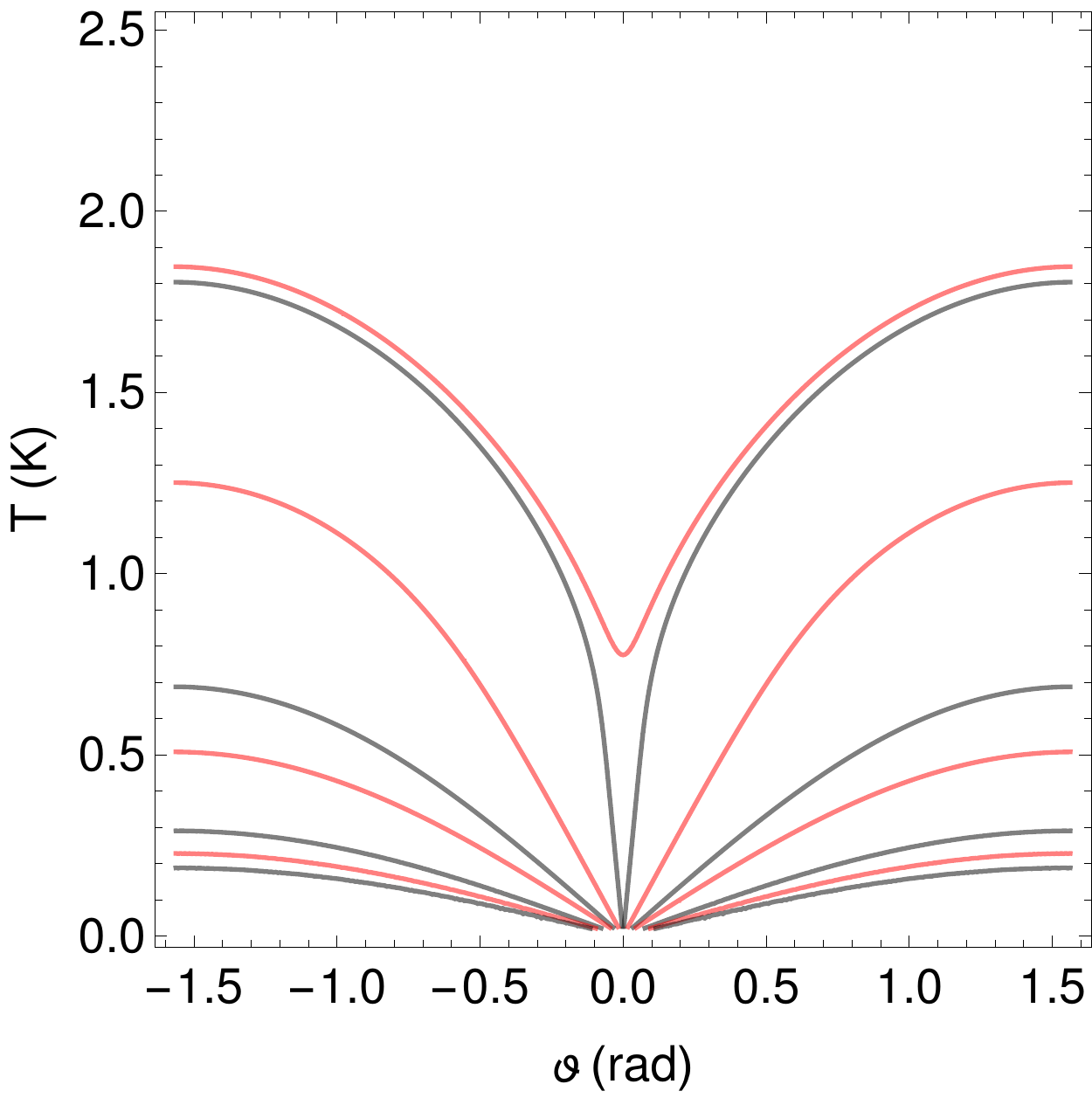}\\
  \includegraphics[width=0.3\textwidth]{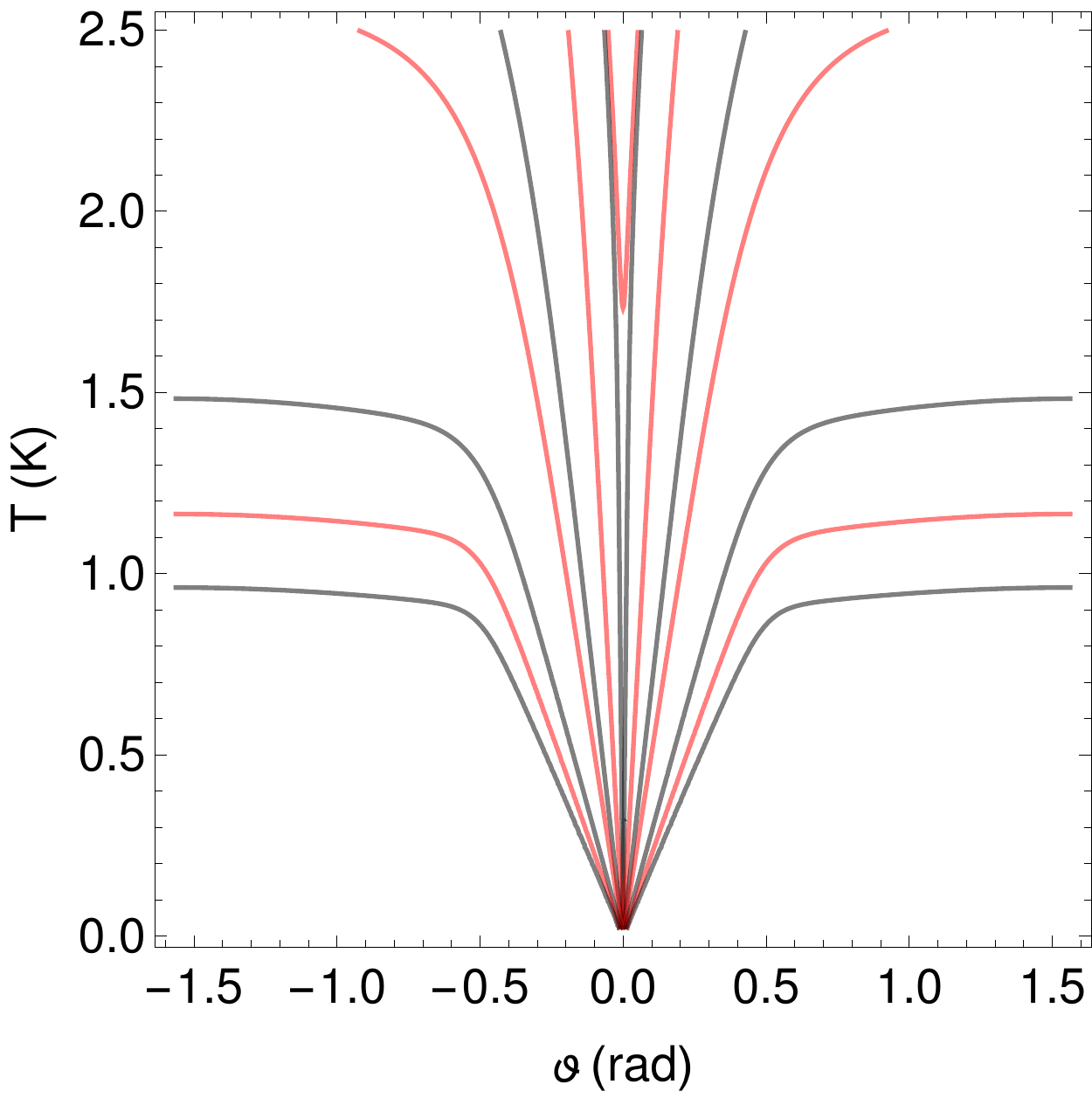}
  &
  \includegraphics[width=0.3\textwidth]{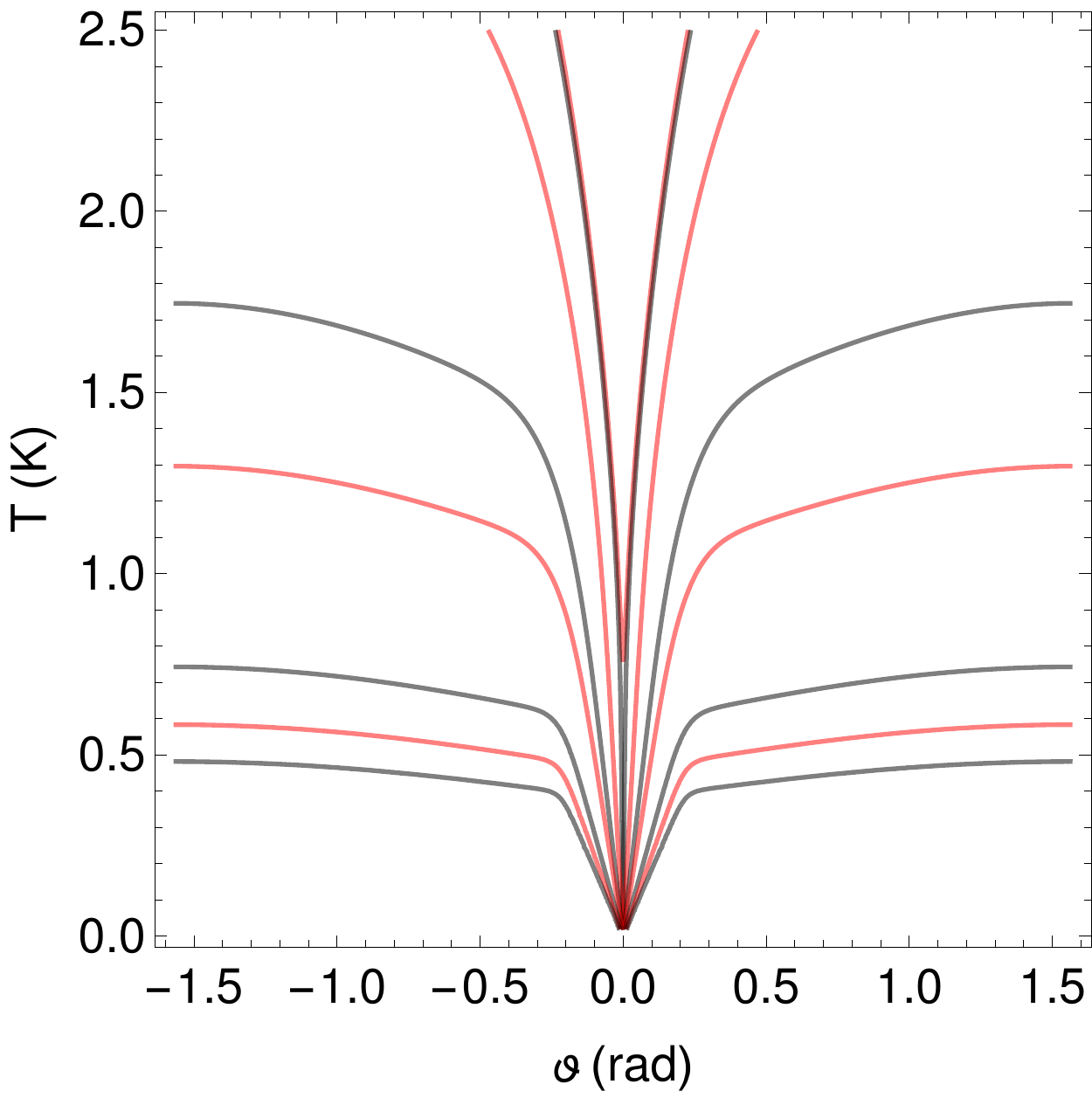}\\
  \includegraphics[width=0.3\textwidth]{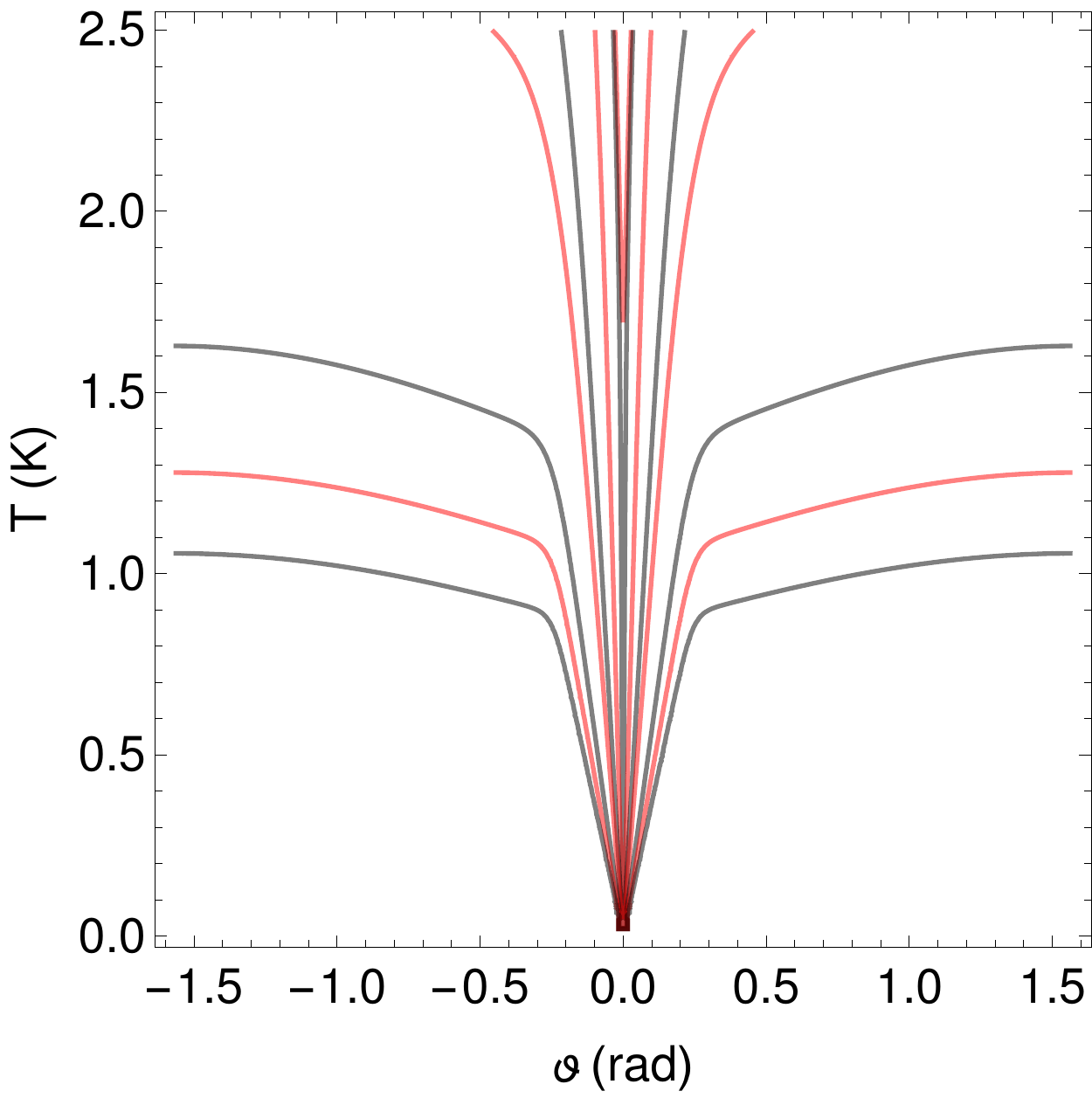}
  &
  \includegraphics[width=0.3\textwidth]{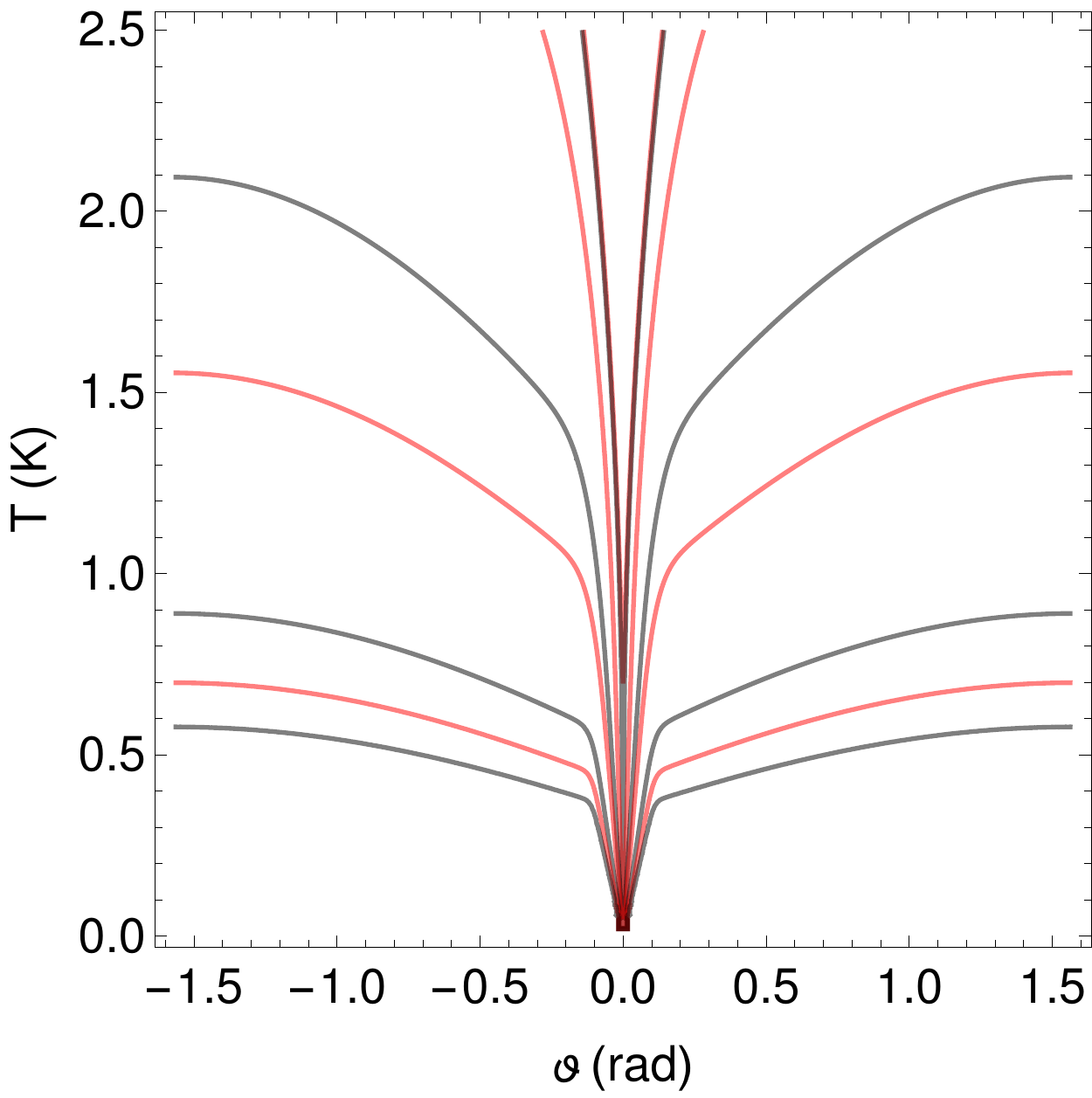}
\end{tabular}
\caption{Isentropes of Mn$_{12}$ (left column) and Fe$_8$ (right
  column) for $B=0.1$~T, $B=1$~T, and $B=2$~T from top to
  bottom. Contours are drawn at entropy values $\mathcal{S}/k_B$ of
  $10^{-5}$, $10^{-4}$, $10^{-3}$, $0.032$, $0.1$, $0.4$,
  $0.68$, and $0.7$.\label{jmmm-amce-f-b}} 
\end{figure}

\section{Quasi-static MCE}
\label{sec-3}

\begin{figure}[ht!]
\centering
\begin{tabular}{ccc}
  \includegraphics[width=0.3\textwidth]{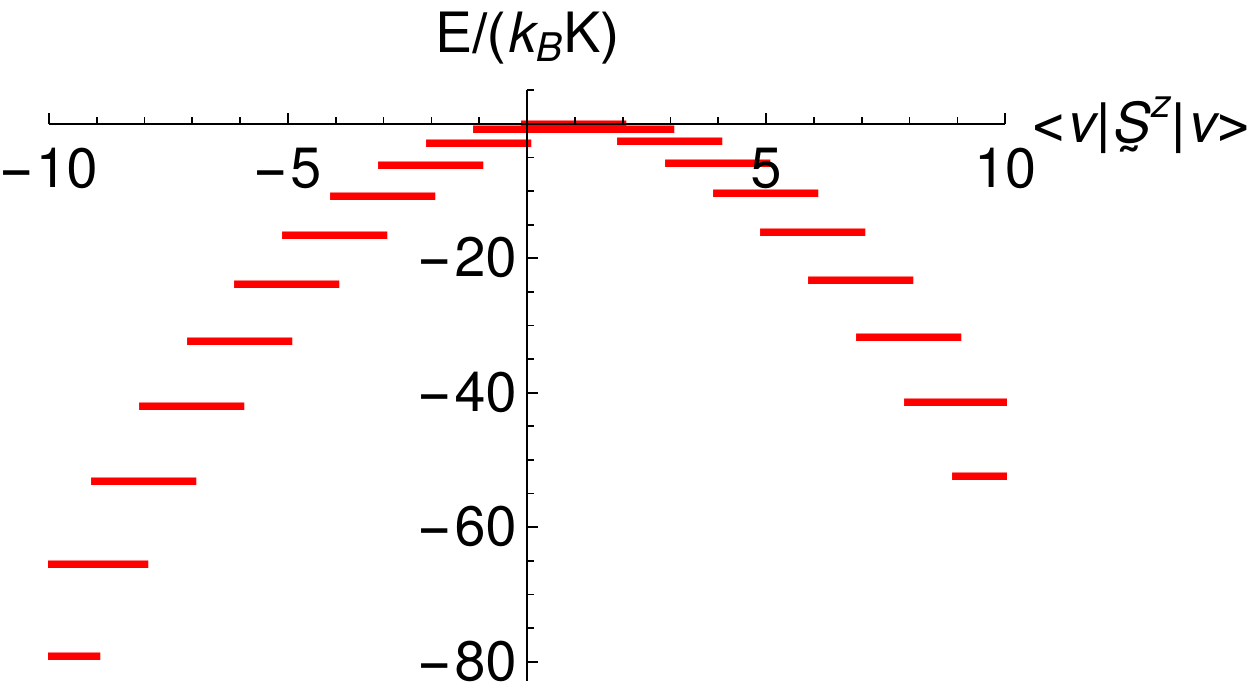}
  &
  \includegraphics[width=0.3\textwidth]{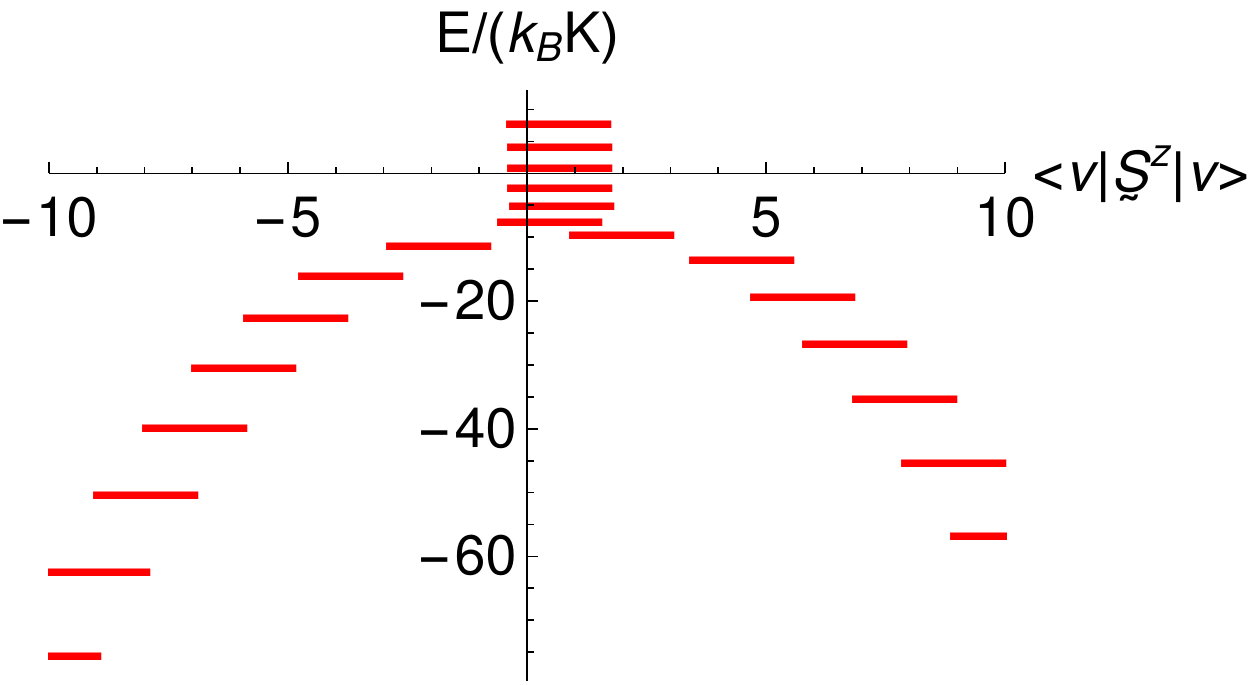}
  &
  \includegraphics[width=0.3\textwidth]{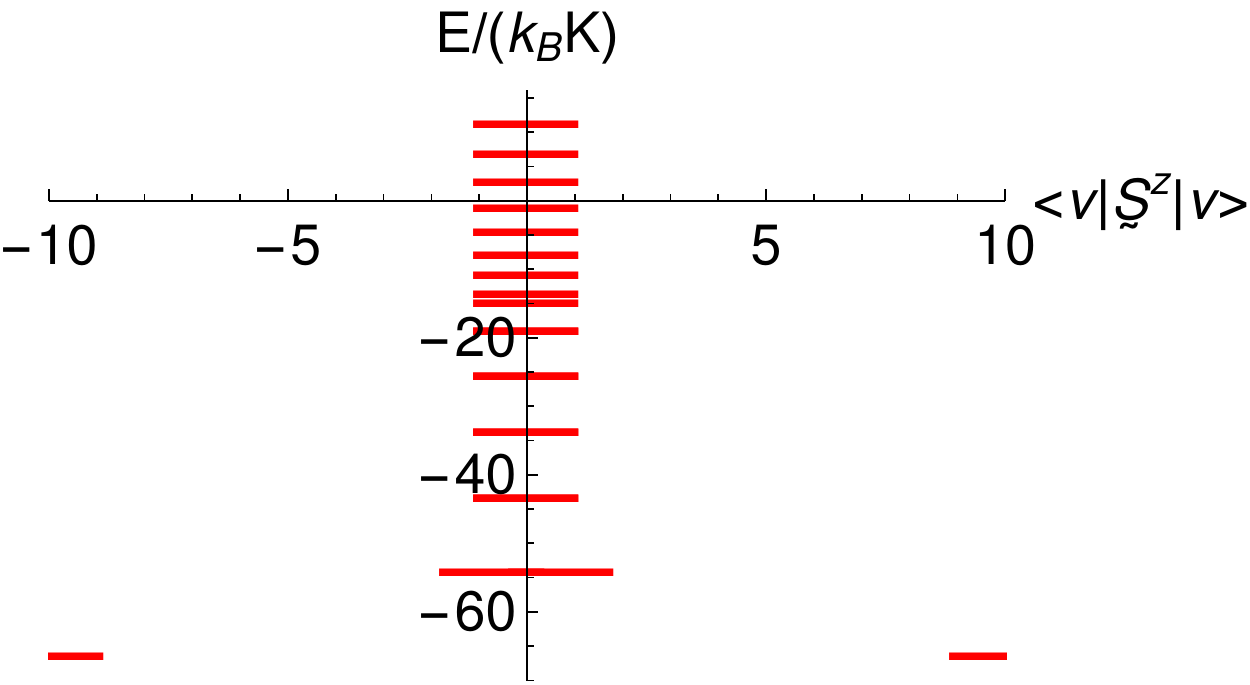}
\end{tabular}
\caption{Energy eigenvalues vs. magnetization along $z$-direction
  for Mn$_{12}$ at $B=1$~T and $\vartheta=\pi/2$ (left),
  $\vartheta=\pi/4$ (middle), and $\vartheta=0$ (right).
\label{jmmm-amce-f-c}} 
\end{figure}

Quasi-static (equilibrium) MCE investigates the thermodynamic
functions \fmref{E-2-5-S} \& \fmref{E-3-5-M} as given by (e.g.)
the canonical ensemble. Of special interest are the isentropes,
i.e. curves of constant entropy, whose slopes are the so-called
cooling rates as well as the isothermal entropy changes -- both
figures of merit for MCE materials. 
Figure \ref{jmmm-amce-f-b} shows the isentropes of Mn$_{12}$
(left column) and Fe$_8$ (right column) for $B=0.1$~T, $B=1$~T,
and $B=2$~T from top to bottom. Since both systems are modeled with
rather similar Hamiltonians, the graphs for these two
SMMs do look very similar. The behavior can be rationalized as
follows. For a given and not too large magnitude of the external
magnetic field 
the energy spectrum resembles a tilted parabola for
$\vartheta=\pi/2$ (l.h.s. of \figref{jmmm-amce-f-c}). In
particular, the ground state is not degenerate. This situation
changes towards $\vartheta=0$ (r.h.s. of
\figref{jmmm-amce-f-c}), where the two ground state levels are
virtually degenerate. This means that all isentropes with
$\mathcal{S}/k_B \leq \log 2$ head towards absolute zero at
$\vartheta=0$. In addition, the top panels of
\figref{jmmm-amce-f-b} display isentropes with
$\mathcal{S}/k_B=0.7>\log 2$ that only exhibit local minima at
$\vartheta=0$. All plots are symmetric about $\vartheta=0$.

The cooling rate (slope of isentropes) assumes very large values
close to $\vartheta=0$. This trend increases with increasing
magnitude of the applied field. Therefore, large
temperature variations should be achievable with only mild
rotations in particular for stronger fields.

\begin{figure}[ht!]
\centering
\begin{tabular}{cc}
  \includegraphics[width=0.4\textwidth]{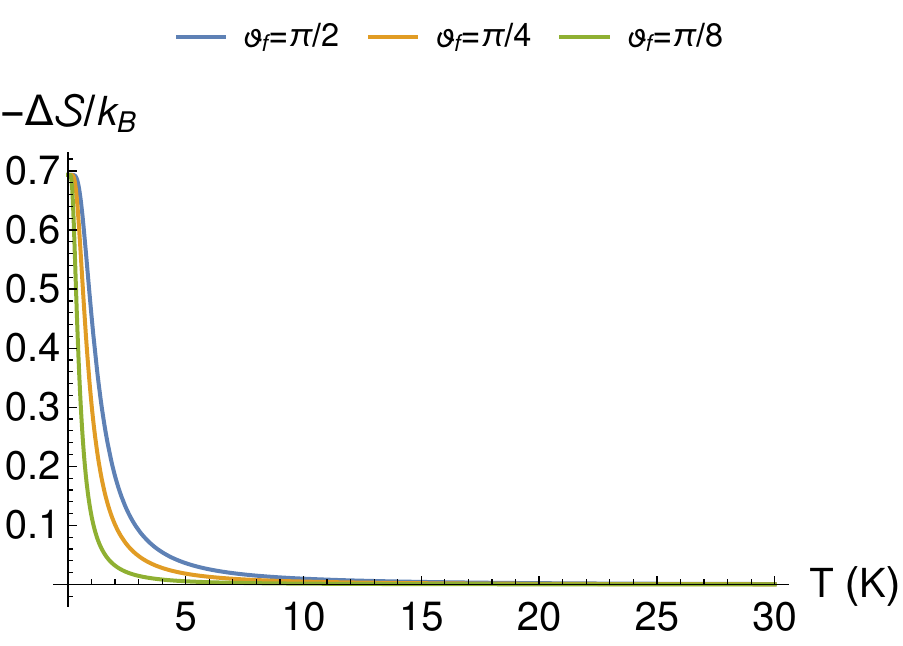}
  &
  \includegraphics[width=0.4\textwidth]{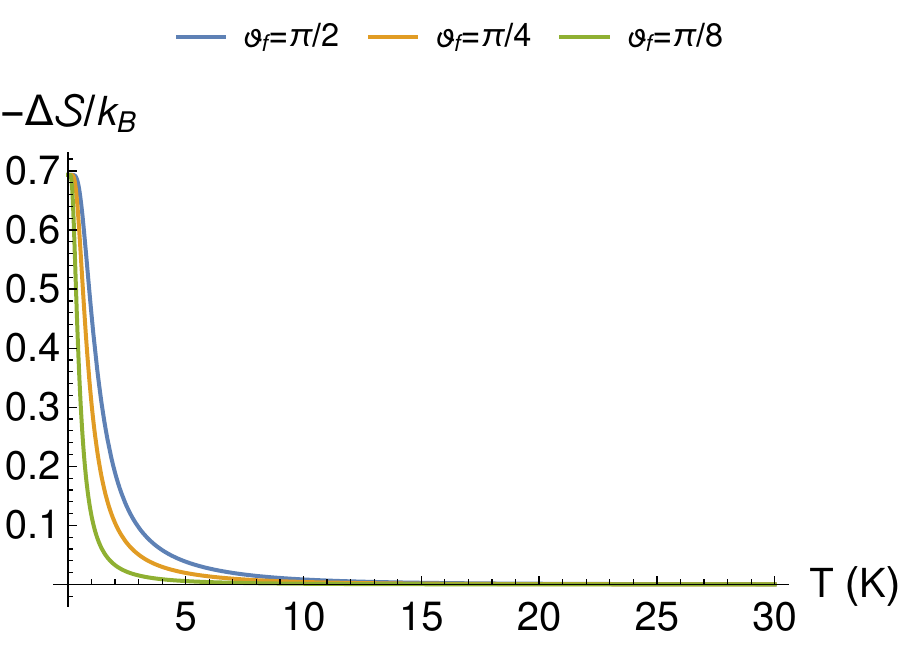}\\
  \includegraphics[width=0.4\textwidth]{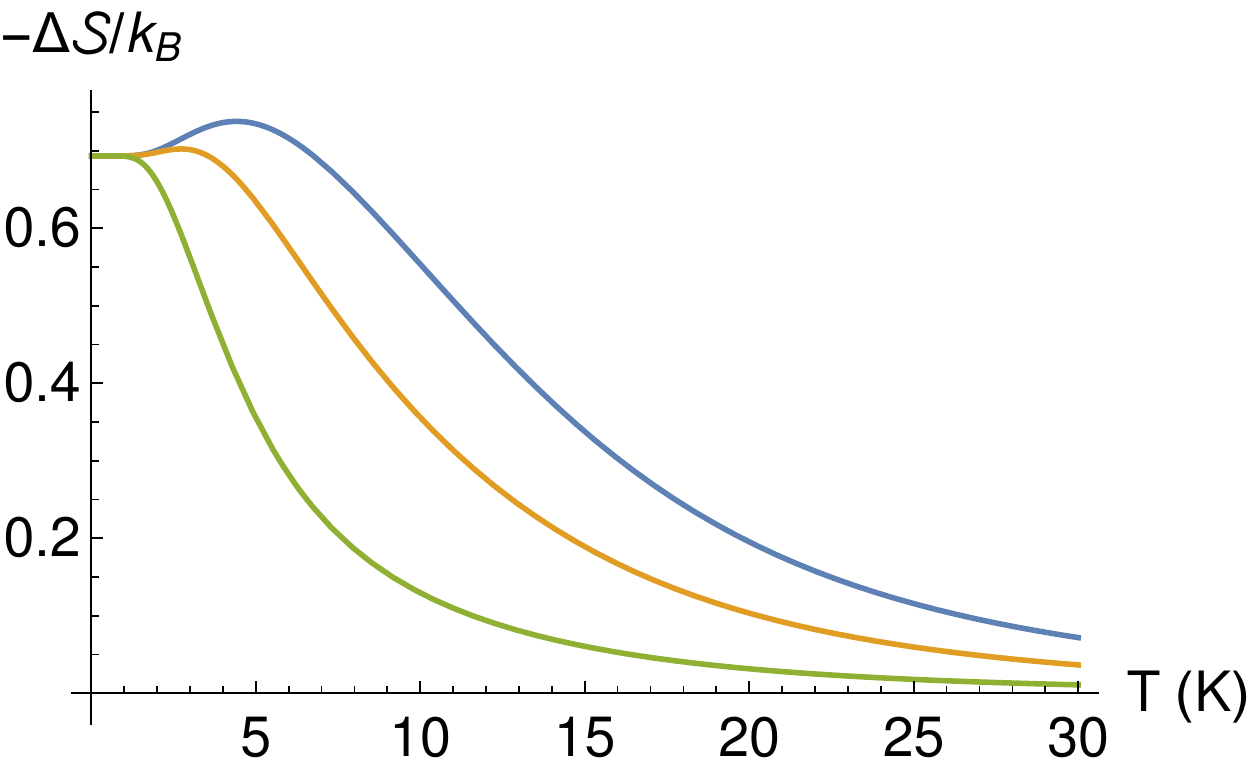}
  &
  \includegraphics[width=0.4\textwidth]{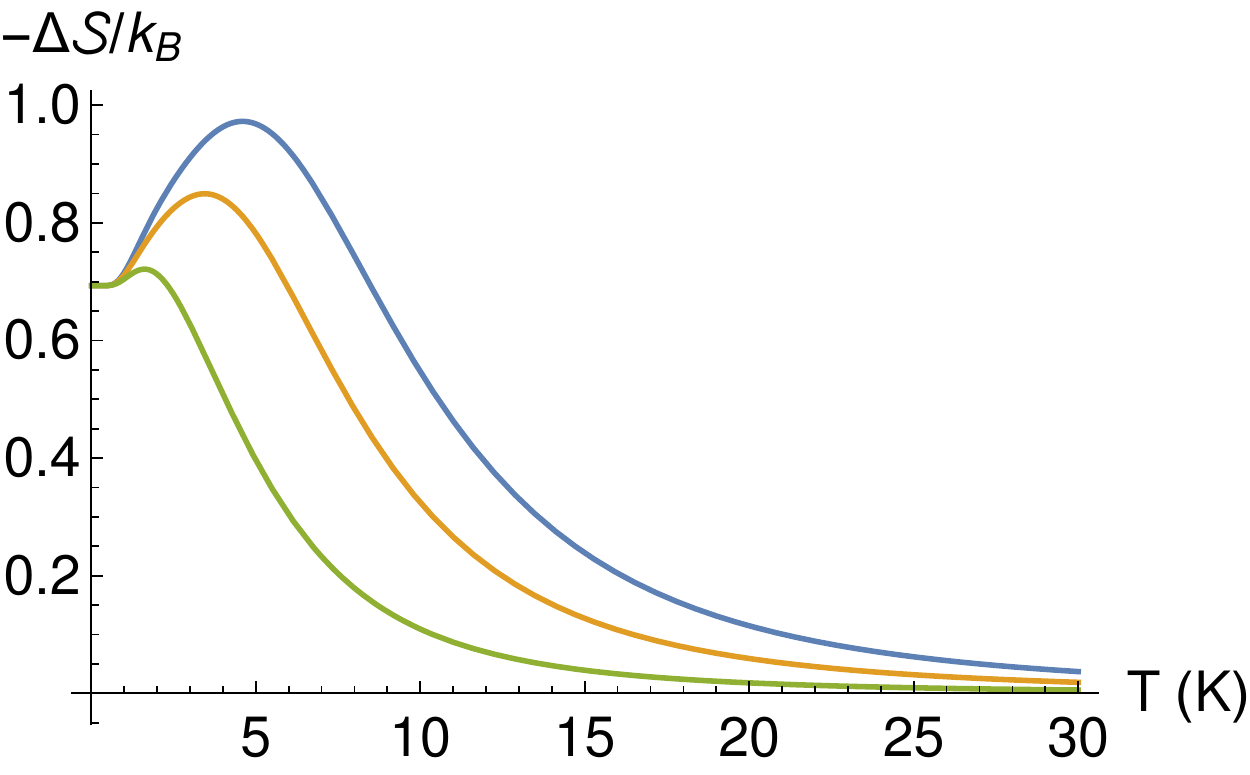}\\
  \includegraphics[width=0.4\textwidth]{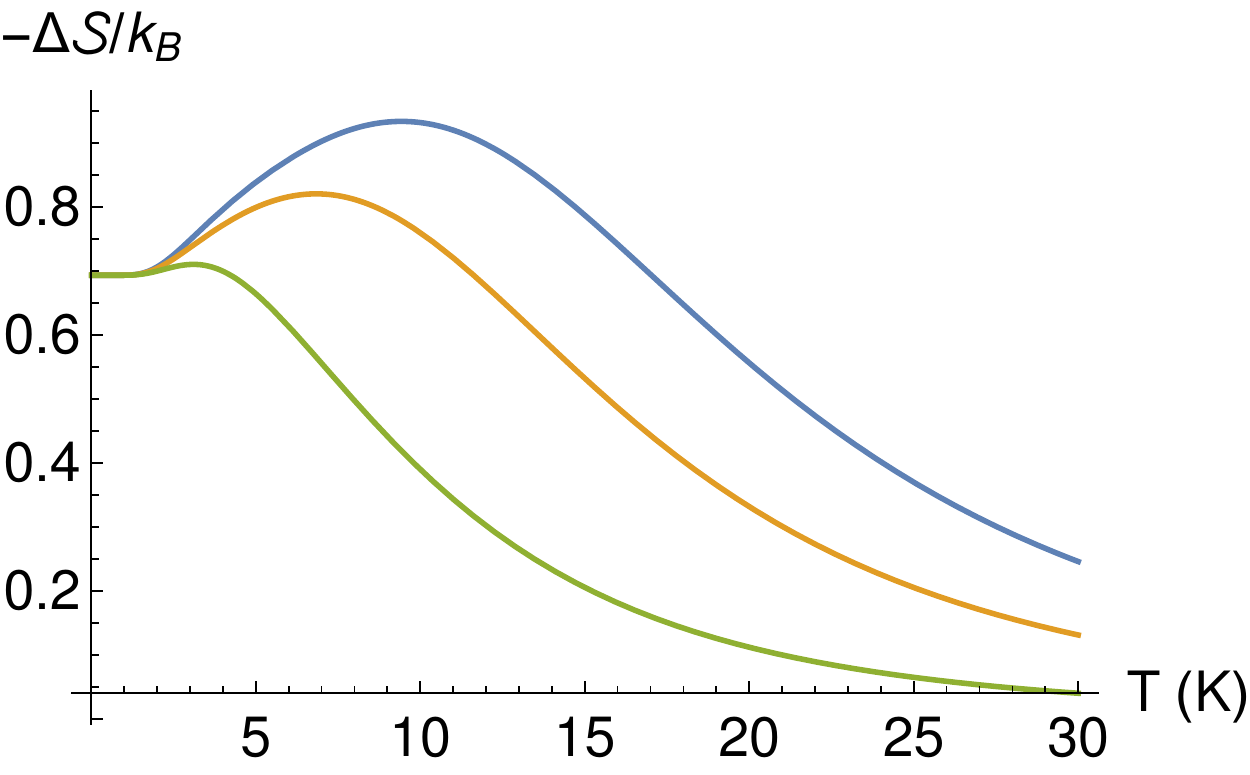}
  &
  \includegraphics[width=0.4\textwidth]{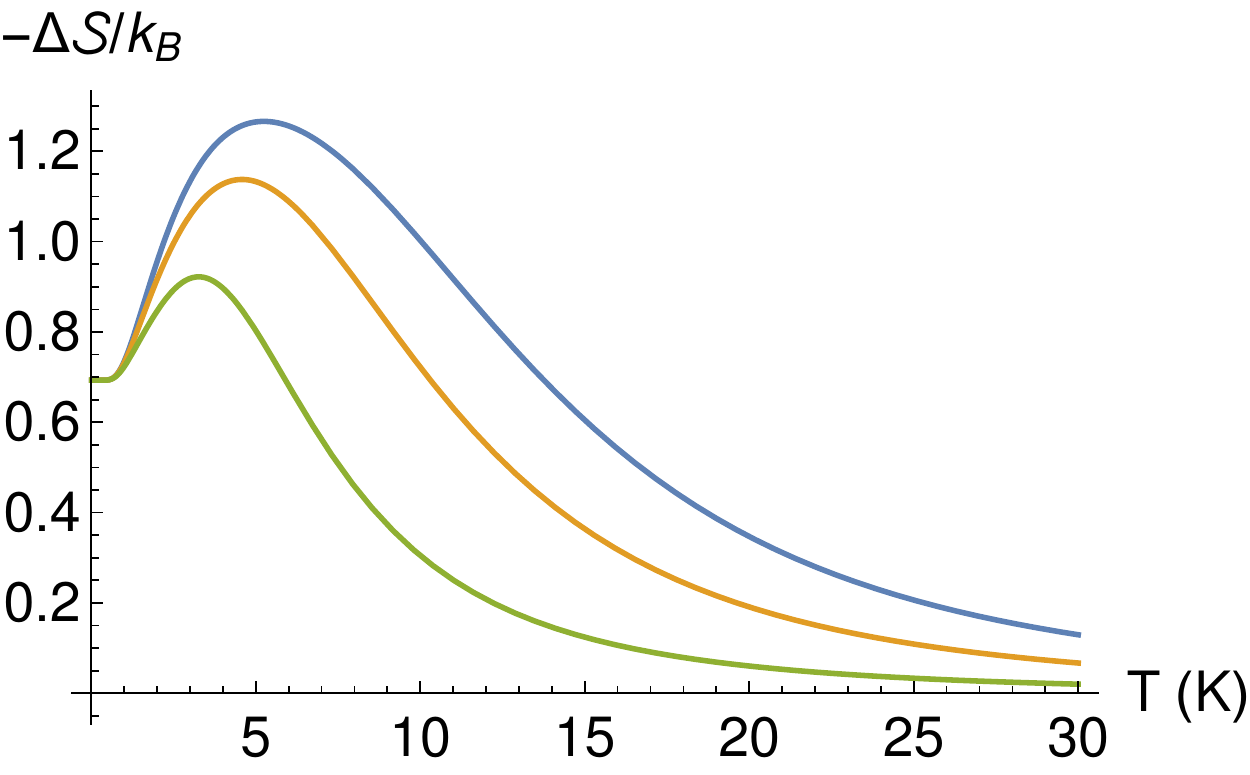}
\end{tabular}
\caption{Isothermal entropy changes of Mn$_{12}$ (left column)
  and Fe$_8$ (right column) for $B=0.1$~T, $B=1$~T, and $B=2$~T
  from top to bottom. The rotation is performed from the initial
angle $\vartheta_i=0$ to the final angles $\vartheta_f$ provided at
the top.\label{jmmm-amce-f-d}} 
\end{figure}

The isothermal entropy change on the other hand is rather
bounded since more than a twofold degeneracy of levels is not
achievable in the physically permitted temperature and field ranges
of the model. This leads to the characteristic curves displayed
in \figref{jmmm-amce-f-d}. Shown is the negative entropy
difference between final and initial orientation, i.e.
\begin{eqnarray}
\label{E-5-1}
-\Delta {\mathcal S}
&=&
-\big(
{\mathcal S}(T,B,\vartheta_f)
-
{\mathcal S}(T,B,\vartheta_i)
\big)
\ .
\end{eqnarray}
The initial angle is always taken as $\vartheta_i=0$. The
colors of the three curves in each panel correspond to the three
chosen final angles $\vartheta_f$ displayed above the panels.

As one notices in all panels of \figref{jmmm-amce-f-d} the
isothermal entropy changes head for $\mathcal{S}/k_B = \log
2\approx 0.68$ at low temperatures. This is a result of the
twofold degeneracy at $\vartheta_i=0$ and the vanishingly small
entropy at all other angles $\vartheta_f\neq 0$. For 
elevated temperatures the entropy change rises a bit since then
also higher lying levels are thermally populated. But due to the
restricted number of levels, which are separated by gaps of the
order of the anisotropy, this effect is small, albeit more
pronounced for stronger external fields. The biggest entropy
changes can be achieved by a rotation of $\Delta\vartheta=\pi/2$
from the direction perpendicular to the easy axis into the
direction of the easy axis.

\section{Realistic Carnot processes}
\label{sec-4}

A discussion of the magnetocaloric properties as in section
\ref{sec-3} or many publications of the field rests on the
assumption of thermal equilibrium, i.e. on idealized
quasi-static processes. However, a realistic cooling experiment
or Carnot process is executed on short time scales of
e.g. minutes \cite{SCM:NC14} or shorter. Whether the system
stays close to equilibrium depends on its typical relaxation
times. In addition, especially for small quantum systems, it is
not granted that the isolated parts of the processes, where no
thermal contact is established, are indeed adiabatic. They may
as well be unitary which is not the same. We do not want to get
into this very complicated discussion and therefore assume that
the isolated steps of our processes are described by a unitary
time evolution. This assumption appears further justified since
we investigate only fast processes in the
following. Investigations of slower processes and processes
other than Carnot are postponed to future investigations.

The Carnot process consists of two isothermal (strokes II and IV)
and two isolated processes (strokes I and III). The time evolution
of the medium, in our case a single Mn$_{12}$ SMM, is modeled via
the time evolution of its density matrix according to \fmref{E-3-3}.
Although the cycle time $\tau_{\text{c}}$ is in reality only limited
by the relaxation during the isothermal strokes, we choose for the
sake of simplicity for all four strokes the same time duration. We
choose $T_{\text{h}}=0.65~\text{K}$ as the temperature of the hot
reservoir and $T_{\text{c}}=0.5~\text{K}$ as the temperature of the
cold reservoir, respectively. For the coupling constant $\lambda$
we choose $\lambda_{\text{h}}=\lambda_{\text{c}}=10^{-1}~\text{ps}^{-1}$.
Smaller values of $\lambda$ simply lead to a shift to lower operating
frequencies $f=\tau_{\text{c}}^{-1}$ and a rescaling of the observed
power. Since we investigate the Carnot process in the realization
as a refrigerator important quantities of interest are the cooling
power $P$ and the efficiency $\epsilon$: 
\begin{eqnarray}
P & = & \frac{Q_{\text{c}}}{\tau_{\text{c}}}\;,\label{E-4-1}\\
\epsilon & = & \frac{Q_{\text{c}}}{W}\;,
\end{eqnarray}
where $Q_{\text{c}}$ is the amount of heat taken from the cold reservoir
and $W$ is the amount of work absorbed by the system during one complete
cycle. 

\begin{figure}[ht!]
\centering %
\begin{tabular}{ccc}
\includegraphics[height=0.3\textwidth]{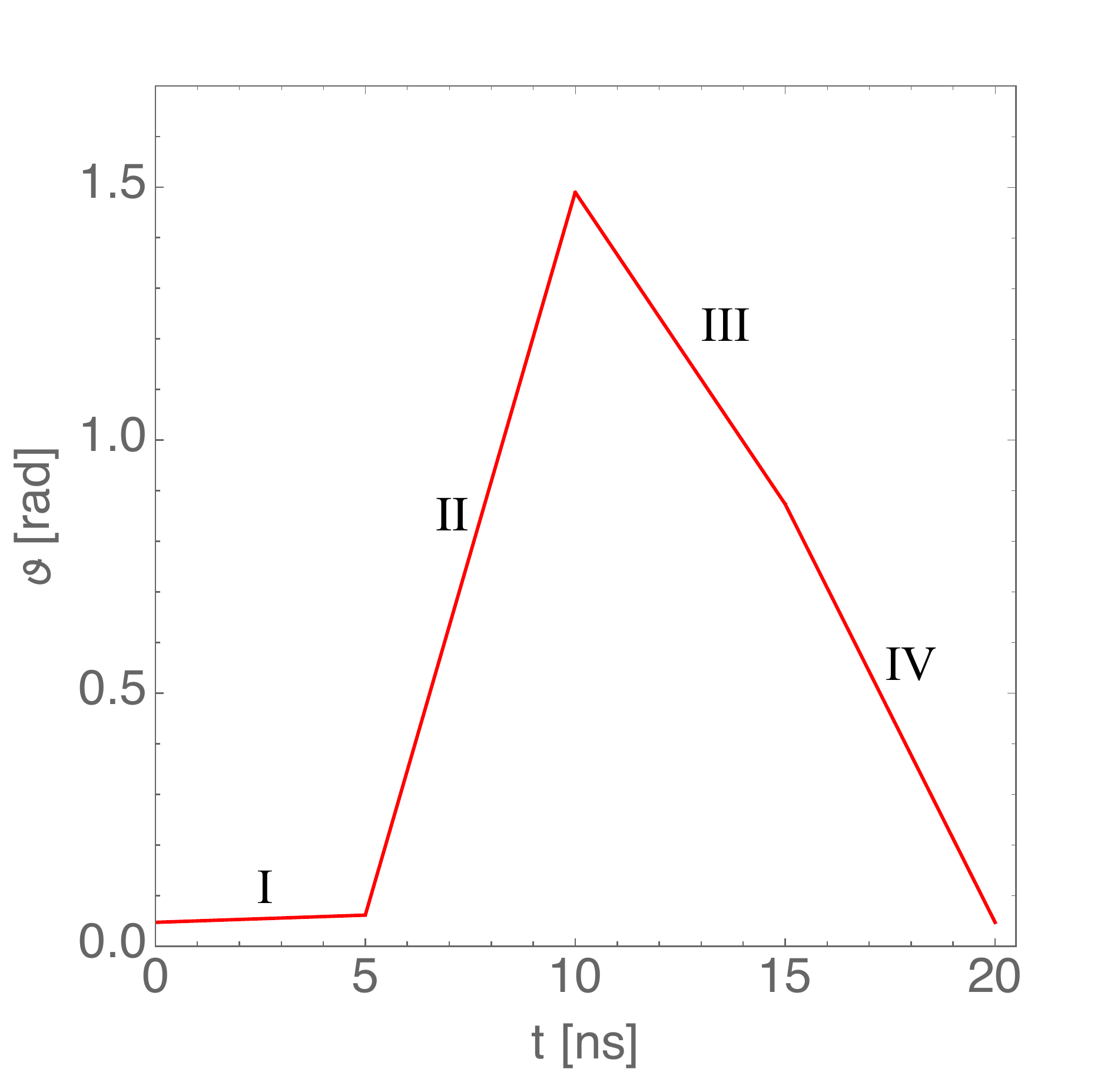}  
& \includegraphics[height=0.3\textwidth]{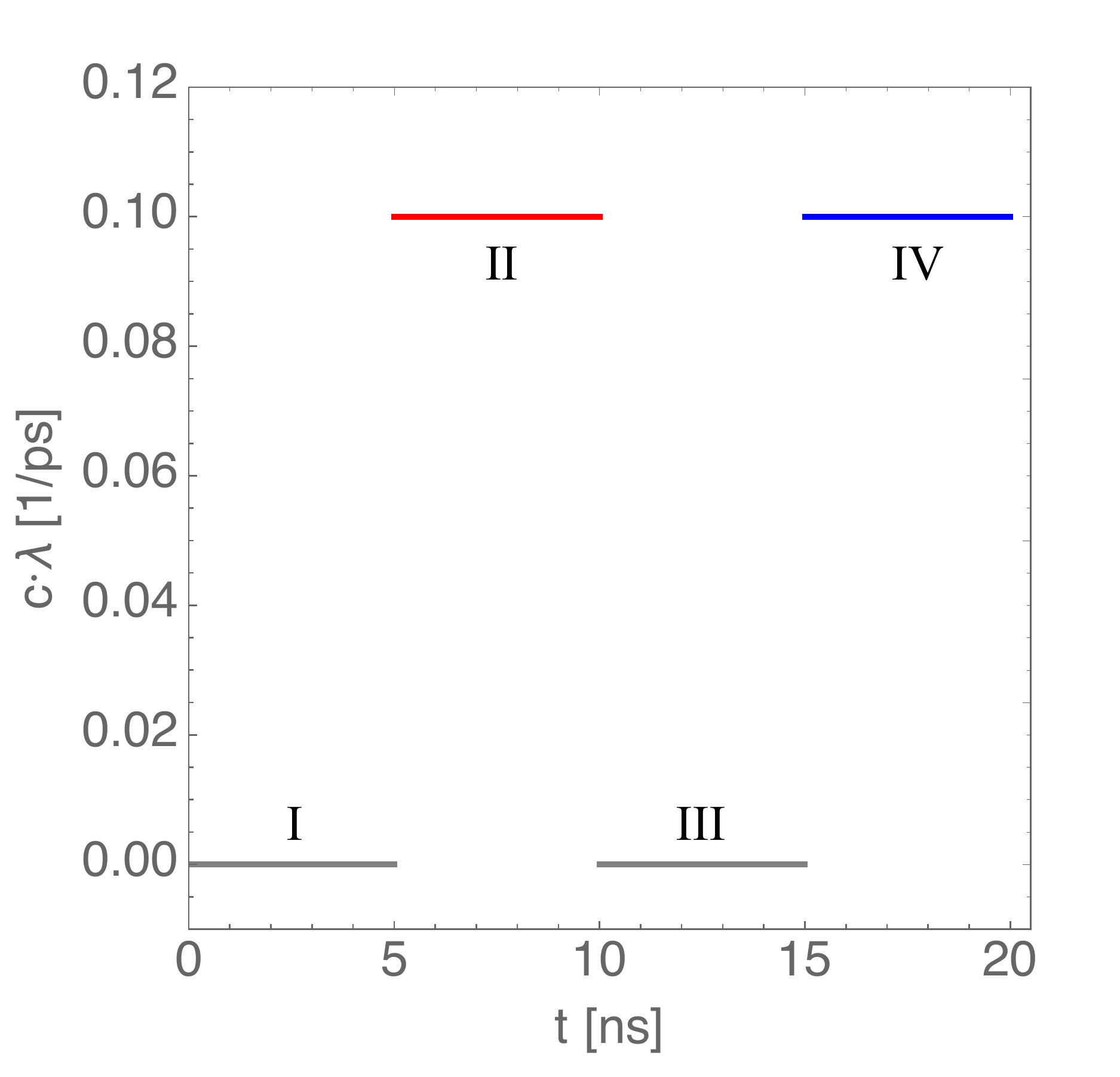}  
&
\includegraphics[height=0.28\textwidth]{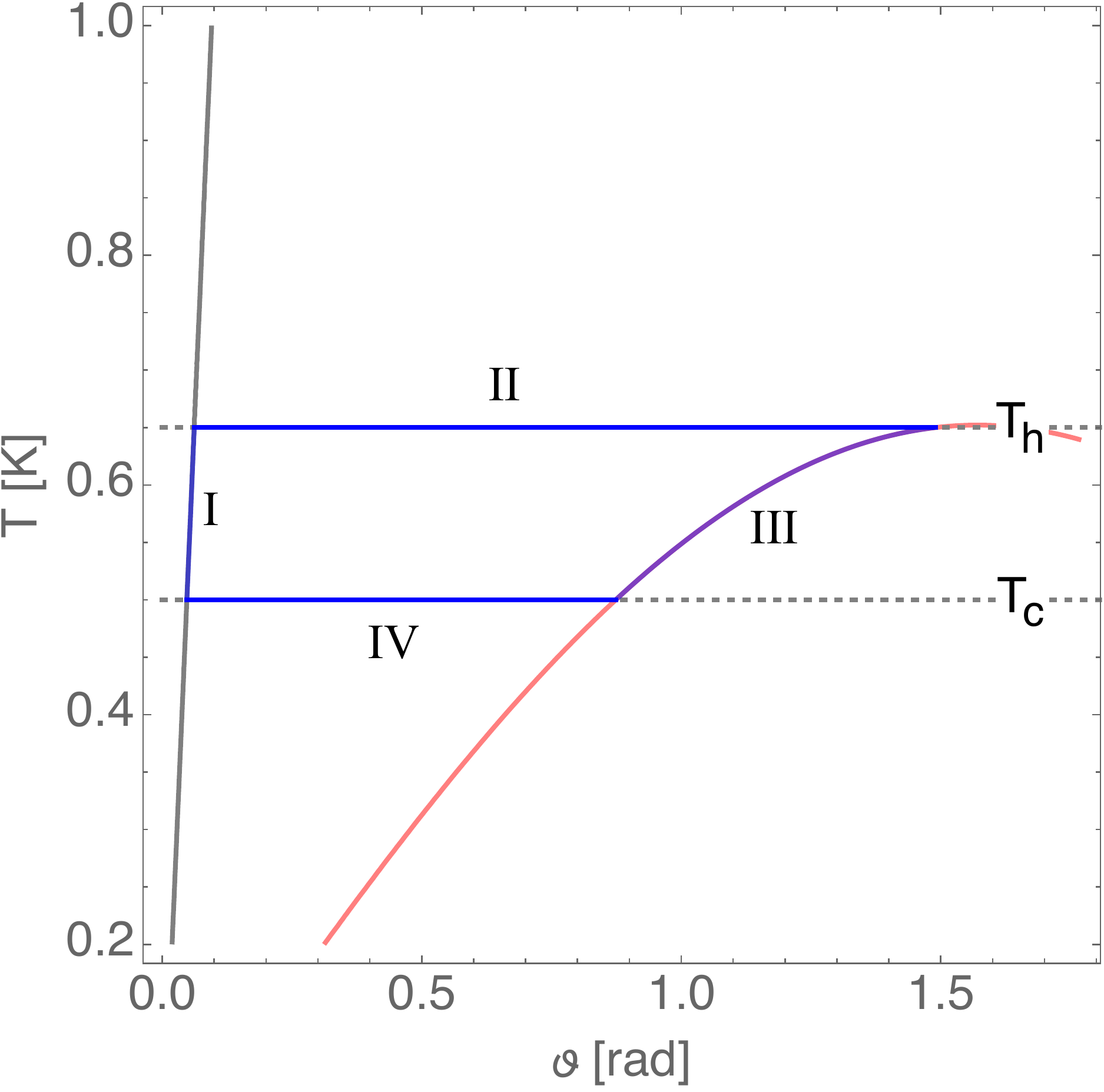} 
\tabularnewline
\end{tabular}
\caption{Exemplaric time dependence of the angle $\vartheta(t)$ (left hand
side) and of the coupling constant $c\cdot\lambda$ (middle). The coupling
to the hot reservoir $\lambda_{\text{h}}$ is shown in red, the coupling
to the cold reservoir $\lambda_{\text{c}}$ is shown in blue. The
right hand side shows an equilibrium, i.e. fully relaxed Carnot
process in the $T$-$\vartheta$-diagram. \label{jmmm-amce-f-4-1}}
\end{figure}

The time dependent angle $\vartheta(t)$ and the behavior of the
coupling constant $c\cdot\lambda$ is exemplarily shown in \figref{jmmm-amce-f-4-1}
for a cycle time of $\tau_{\text{c}}=20~\text{ns}$. Figure \figref{jmmm-amce-f-4-1}
also shows the process in the corresponding $T$-$\vartheta$-diagram.

During the beginning of stroke I the system is in thermal equilibrium
with the cold heat reservoir at temperature $T_{\text{c}}$. The system
is then decoupled from the heat reservoir and the angle $\vartheta$
is changed with constant angular velocity from $\vartheta_{0}$ to $\vartheta_{1}$
(compare stroke I in \figref{jmmm-amce-f-4-1}). Since the system
evolves isolated during this stroke there is no heat exchanged with
the reservoirs. The work can therefore be calculated via \fmref{E-3-7}.

During stroke II the coupling to the hot heat reservoir is switched on
while the angle $\vartheta$ is further increased to $\vartheta_{2}$
with a constant (but different to the previous step) velocity (compare
stroke II in \figref{jmmm-amce-f-4-1}). The system relaxes during
this stroke towards thermal equilibrium with the hot heat
reservoir, but depending on the time of contact with the bath,
equilibrium is not necessarily reached.  Since this stroke is isothermal the
work must be calculated via \fmref{E-3-6}. 
The amount of heat $Q_{\text{h}}$ exchanged with the hot heat reservoir
can then be calculated via \fmref{E-3-7}.

For stroke III the system is again decoupled from the heat reservoir
and the angle $\vartheta$ is decreased with another constant velocity
from $\vartheta_{2}$ to $\vartheta_{3}$ (compare stroke III in \figref{jmmm-amce-f-4-1}).
Because this stroke is again isolated and there is again no heat exchange
with any of the heat reservoirs the work can be calculated directly
from \fmref{E-3-7}. 

During the last stroke IV the system is coupled to the cold heat reservoir
at temperature $T_{\text{c}}$ while the angle $\vartheta$ is decreased
with another constant velocity until the initial angle $\vartheta_{0}$
is reached and the cycle is complete (compare stroke IV in
\figref{jmmm-amce-f-4-1}). The system evolves towards thermal
equilibrium with the cold heat reservoir as
much as possible during contact time. Since this stroke is again isothermal
the work must be calculated via \fmref{E-3-6}. The amount of heat
$Q_{\text{c}}$ exchanged with the cold heat reservoir can then be
calculated from \fmref{E-3-7}.

The observables presented in the following parts are evaluated
after the system has been driven through sufficiently many
cycles in order to reach a steady state.

\subsection{Dependence of power and operating frequencies on the amplitude of
$\vec{B}(t)$}

\begin{figure}[ht!]
\centering 
\includegraphics[width=0.45\textwidth]{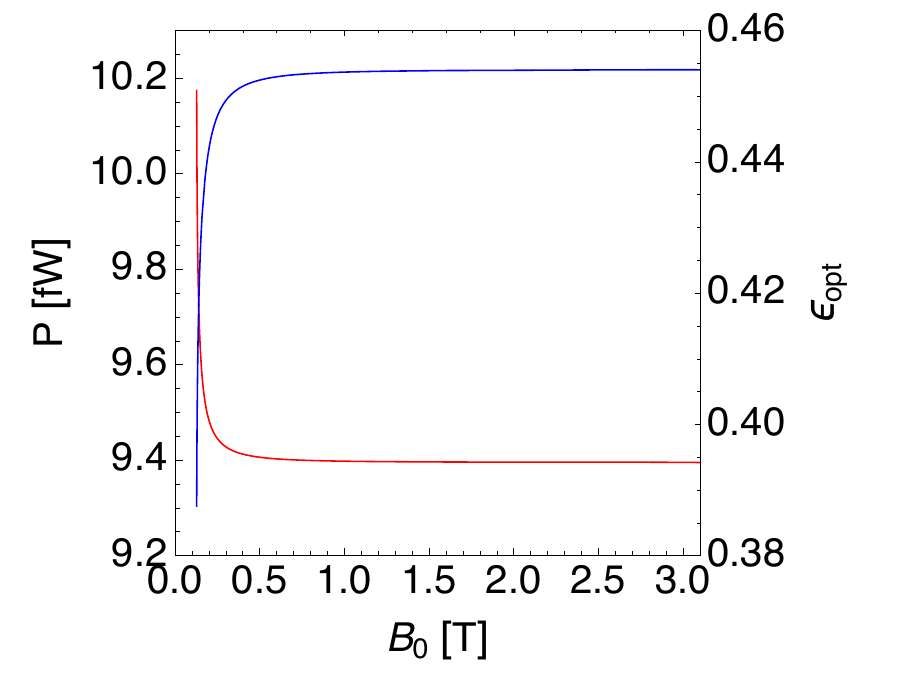}\quad
\includegraphics[width=0.45\textwidth]{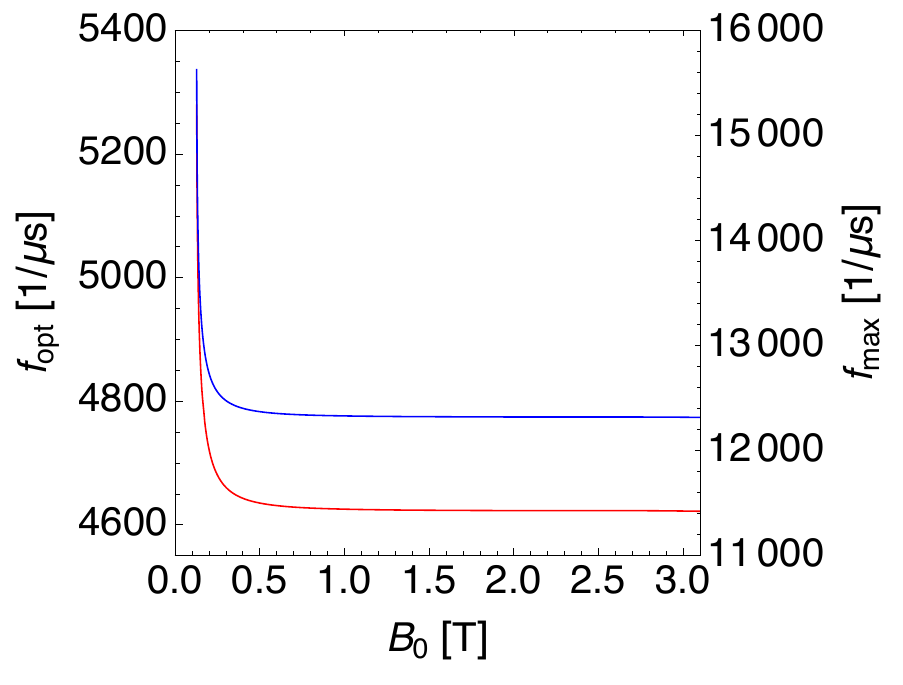}
\caption{Dependence of the maximum cooling power $P$ and
  corresponding efficiency 
$\epsilon_{\text{opt}}$ (left hand side) and the optimal and
  maximal operating 
frequency $f$ (right hand side) on the amplitude of the magnetic
field $B_{0}$. The curves belonging to the left axis are shown in
red, the curves belonging to the right axis are shown in
blue. Different scales are used. \label{jmmm-amce-f-4-2}} 
\end{figure}

At first we investigate the influence of the amplitude $B_{0}$ of
the magnetic field $\vec{B}(t)$ on the maximum cooling power $P$,
the optimal operating frequency $f_{\text{opt}}$, the efficiency
$\epsilon_{\text{opt}}$ as well as on the maximum operating frequency $f_{\text{max}}$.
Here the optimal operating frequency $f_{\text{opt}}$ denotes the
operating frequency and $\epsilon_{\text{opt}}$ the efficiency at maximum cooling
power. The maximum operating frequency $f_{\text{max}}$ is the maximal
frequency for which the Carnot cycle works as a refrigerator delivering
heat from the cold heat reservoir to the hot one by consuming work.
The results of our simulations are shown in \figref{jmmm-amce-f-4-2}.
The angles $\vartheta_{0}$ to $\vartheta_{3}$ are chosen such that
the process always operates between the two isentropes 
$\mathcal{S}_{\text{max}}/k_{B}=0.68$
and $\mathcal{S}_{\text{min}}/k_{B}=0.032$ and therefore with a fixed
$\Delta\mathcal{S}/k_{B}=0.648$.

The minimal possible amplitude $B_{0}$ that can satisfy
$\mathcal{S}_{\text{min}}$ and 
$\mathcal{S}_{\text{max}}$ at the given temperatures of the heat
reservoirs is $B_{0}=0.128~\text{T}$. As one can deduce from
\figref{jmmm-amce-f-4-2}, $P$, $f_{\text{opt}}$ and
$f_{\text{max}}$ 
are maximal for this amplitude. Only the efficiency at maximum power
$\epsilon_{\text{opt}}$ is minimal. When one increases the amplitude $B_{0}$,
$P$ decreases by $7.66\%$ until the amplitude $B_{0}$ reaches a threshold
value of about $0.5~\text{T}$. The optimal operating frequency $f_{\text{opt}}$
decreases in the same time by $12.45\%$ and $f_{\text{max}}$ decreases
by even $21.17\%$. The efficiency on the other hand increases by
$17.15\%$. For larger values of $B_{0}$ all observed quantities
become independent of $B_{0}$. 

For the hot and cold temperatures $T_{\text{c}}$ and
$T_{\text{h}}$, respectively, chosen in our example, one also
deduces from \figref{jmmm-amce-f-b} that with increasing field
strength $B_{0}$ the maximum rotation angle
decreases. Therefore, for large amplitudes $B_{0}$ only very
small rotations are necessary.

\begin{figure}[ht!]
\centering \includegraphics[width=0.4\textwidth]{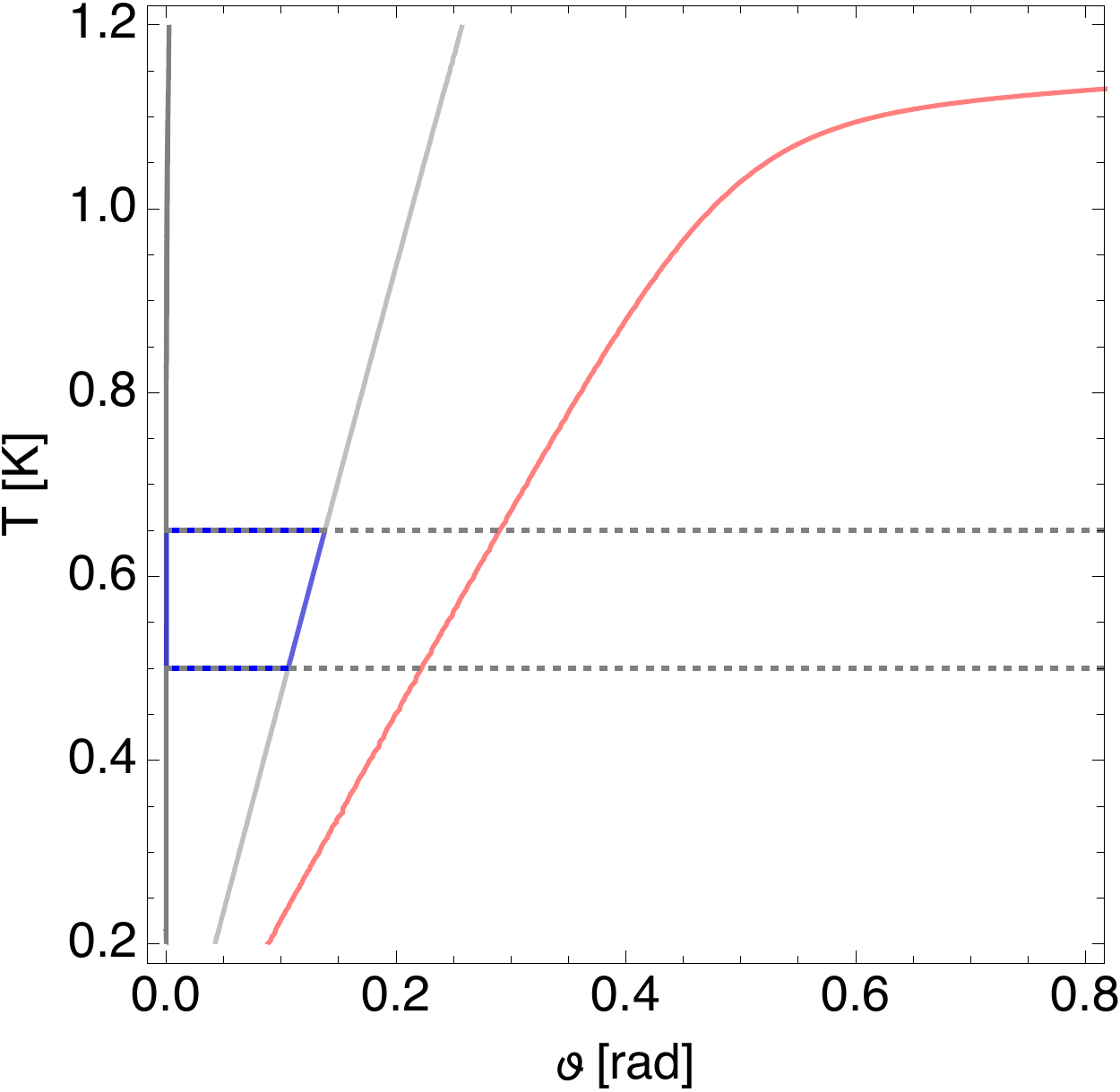}
\caption{Exemplaric depiction of the Carnot cycle in the $T$-$\vartheta$-diagram.
The isentrope $\mathcal{S}_{\text{max}}$ (black line) is fixed at
$\mathcal{S}/k_{B}=0.6931$ while the isentrope $\mathcal{S}_{\text{min}}$ (gray
line) is varied between black line and red curve to achieve
different $\Delta\mathcal{S}$. The temperatures 
of the two heat reservoirs are marked by dotted gray lines. A Carnot
cycle is depicted in blue.
\label{jmmm-amce-f-4-3}}
\end{figure}

\subsection{Dependence of power and operating frequencies on $\Delta\mathcal{S}$}

The quasi-static solution of the Carnot process yields a linear dependence
between the heat extracted from the cold heat reservoir $\Delta Q_{\text{c}}$
and the entropy difference $\Delta\mathcal{S}$ between the two isentropes
$\mathcal{S}_{\text{min}}$ and $\mathcal{S}_{\text{max}}$ of the
process: 
\begin{eqnarray}
\Delta Q_{\text{c}} & = & T_{\text{c}}\cdot\Delta\mathcal{S}\;.\label{E-4-2}
\end{eqnarray}
Thus a large value of $\Delta\mathcal{S}$ is intended to
maximize the cooling per cycle. To investigate if this still holds for the dynamic process
we investigate again the maximum cooling power $P$ and the corresponding
efficiency at maximum cooling power $\epsilon_{\text{opt}}$ as well as the operating
frequencies $f_{\text{opt}}$ and $f_{\text{max}}$. The amplitude
$B_{0}$ of the applied magnetic field is fixed at $B_{0}=1~\text{T}$.
We also fix the isentrope $\mathcal{S}_{\text{max}}$ to a value that
is very close to the maximal possible value. We use $\mathcal{S}_{\text{max}}=0.6931~k_{B}$.
The other isentrope $\mathcal{S}_{\text{min}}$ is varied to achieve
different $\Delta\mathcal{S}$. This is exemplarily shown in \figref{jmmm-amce-f-4-3}. 

\begin{figure}[ht!]
\centering 
\includegraphics[width=0.45\textwidth]{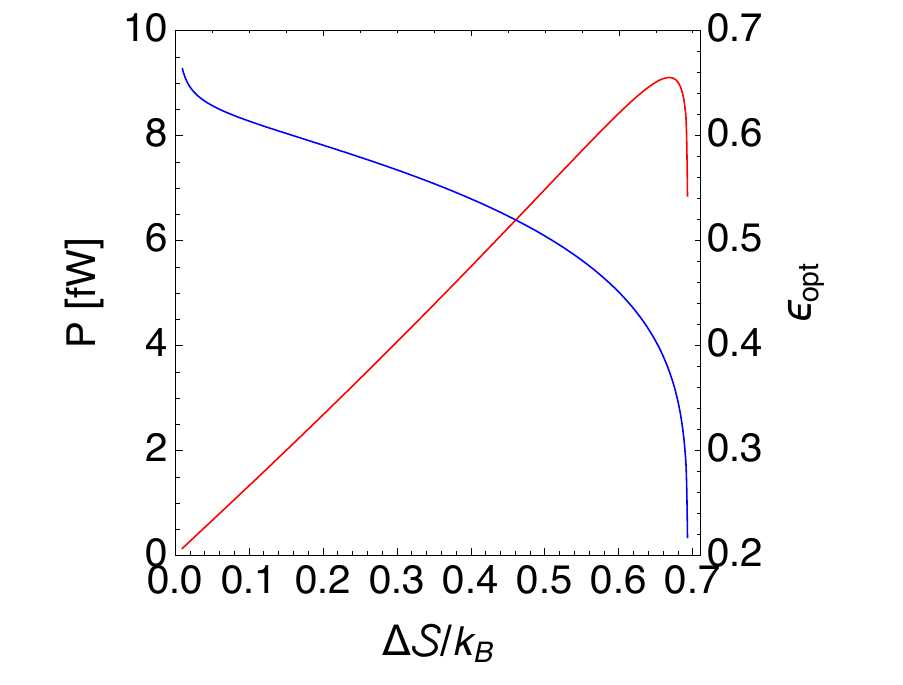}\quad
\includegraphics[width=0.45\textwidth]{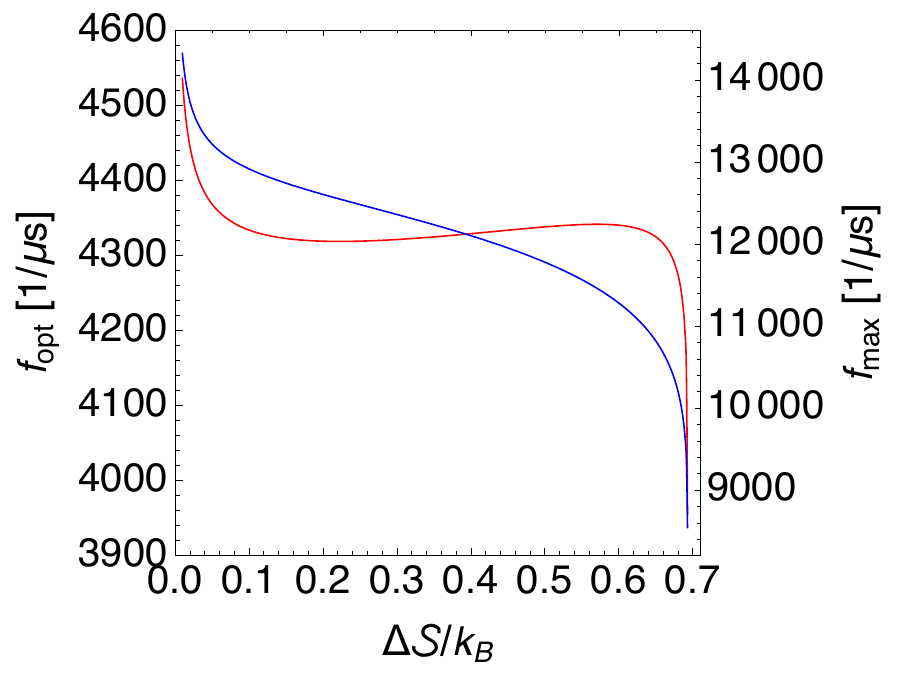}
\caption{Dependence of the maximum cooling power $P$ and
  corresponding efficiency 
$\epsilon$ (left hand side) and the optimal and maximal operating
frequency $f$ (right hand side) on the value of $\Delta\mathcal{S}$.
The curves belonging to the left axis are shown in red, the curves
belonging to the right axis are shown in blue. Different scales
are used. \label{jmmm-amce-f-4-4}}
\end{figure}

The results of our simulations are shown in \figref{jmmm-amce-f-4-4}.
As one can see from the left hand side of \figref{jmmm-amce-f-4-4} the maximum
cooling power $P$ grows almost linearly with $\Delta\mathcal{S}$
(red curve).
But there is a significant loss of $P$ when $\Delta\mathcal{S}$
gets larger than $\Delta\mathcal{S}_{\text{opt}}=0.668~k_{B}$ (that
is $96.44\%$ of the maximum value of $\Delta\mathcal{S}$). This
loss is about almost $25\%$ when the cycle is operated at maximum
$\Delta\mathcal{S}$ instead of $\Delta\mathcal{S}_{\text{opt}}$.
In contrast to the quasi-static case the largest $\Delta\mathcal{S}$
does not yield the maximum performance. 

The efficiency $\epsilon_{\text{opt}}$ 
at maximum power decreases
monotonically with growing $\Delta\mathcal{S}$ (blue curve in
\figref{jmmm-amce-f-4-4} l.h.s.), and the slope is larger for
small as well as large values of $\Delta\mathcal{S}$. The same is true for the maximum
operating frequency $f_{\text{max}}$ (compare blue curve in
\figref{jmmm-amce-f-4-4} r.h.s.). The optimal operating
frequency $f_{\text{opt}}$, red curve in \figref{jmmm-amce-f-4-4} r.h.s.,
behaves differently, since it has a local minimum at $\Delta\mathcal{S}=0.222~k_{B}$
($32.1\%$ of the maximum $\Delta\mathcal{S}$) and a local maximum
at $\Delta\mathcal{S}=0.570~k_{B}$ ($82.36\%$ of the maximum
$\Delta\mathcal{S}$).

\section{Summary and outlook}
\label{sec-5}

In this article we report investigations of the rotational
magnetocaloric effect using single molecule magnets. We can
conclude that the effect is present and may be used in cases
where quick field changes, that are possible using mechanical
rotations, are necessary. The isothermal entropy change, on the
other hand, is limited since degeneracies larger than two do not
arise and thus the entropy does not grow much above
$\mathcal{S}/k_B \approx \log 2$. 

A description of the Carnot process as a realistic
time-dependent non-equilibrium process -- using a simplified
dynamics -- reveals that for SMMs a threshold field amplitude
exists above which the characteristic figures do not change any
more. In addition and in contrast to the quasi-static case the
largest $\Delta\mathcal{S}$ does not yield the maximum
performance.  Instead the maximum cooling power is achieved with
an optimal value for $\Delta\mathcal{S}$ of only about 96~\% of
the maximum possible value. 

In future investigations slow processes shall be studied, where
the interaction with the heat bath includes the relevant phonon
degrees of freedom in order to take the phonon bottleneck into
account \cite{CWM:PRL00,SCL:PRL05,Gar:PRB07,CGS:PRL08,BeS:JMMM17}.

\section*{Acknowledgment}

The authors thank Wolfgang Wernsdorfer for useful
discussions. Funding by the Deutsche Forschungsgemeinschaft (DFG
SCHN 615/23-1) is thankfully acknowledged.


\end{document}